\documentclass[11pt]{article}
\usepackage{amssymb,amsfonts}
\usepackage[pdftex]{graphicx}
\DeclareGraphicsExtensions{.pdf}  \textheight24.5cm
\begin{document}

\renewcommand{\thefootnote}{\fnsymbol{footnote}}

\centerline{\Large \bf The Euler and Navier--Stokes equations revisited} 
\vspace*{3mm} \centerline{Peter Stubbe\footnote[1]{retired from Max--Planck--Institut f\"ur Sonnensystemforschung, 37077 G\"ottingen, Germany. Contact by e-mail: peter-stubbe@t-online.de }} 
 
\renewcommand{\thefootnote}{\arabic{footnote}}

\vspace*{.5cm}{\small \begin{quote} The present paper is motivated by recent mathematical work on the incompressible Euler and Navier--Stokes equations, partly having physically problematic results and unrealistic expectations. The Euler and Navier--Stokes equations are rederived here from the roots, starting at the kinetic equation for the distribution function in phase space. The derivation shows that the Euler and Navier--Stokes equations are valid only if the fluid under consideration is an ideal gas, and if deviations from equilibrium are small in a defined sense, thereby excluding fully nonlinear solutions. Furthermore, the derivation shows that the Euler and Navier--Stokes equations are unseparably coupled with an appertaining equation for the temperature, whereby, in conjunction with the continuity equation, a closed system of transport equations is set up which leaves no room for any additional equation, with the consequence that the frequently used incompressibility condition $\nabla\cdot{\bf v}=0$ can, at best, be applied to simplify these transport equations, but not to supersede any of them. \end{quote} }

\vspace*{.5cm}{\large \bf 1. Introduction}

The Euler and Navier--Stokes equations belong to the oldest equations in physics. They have been the subjects of innumerable publications
and are widespread over the textbook literature. It may appear, therefore, that nothing new could be written about these equations.

The reason for a renewed interest in these equations has come from the mathematical side, primarily with regard to the Euler equation. A central question in these mathematical studies has been whether the energy of the system is conserved by the soltutions. Scheffer [1] and Shnirelman [2] showed the existence of solutions which seem to indicate that kinetic energy is created out of nothing. Other investigations (e.g., Constantin et al.~[3], Cheskidov et al.~[4], De Lellis and Szekelyhidi [5]), following up Onsager's conjecture, deal with just the opposite, the dissipation of energy without having a loss term in the equations.

From a physical point of view these results are unacceptable. How should simple hydrodynamics have the potential to break a most fundamental physical principle, the conservation of energy in a closed system? But then, how could it be that mathematically correct solutions of physical equations should be physically invalid? The origin will have to be found in the equations, and in a possible violation of the applicability limits of these equations.  

It is tacitly assumed in these mathematical treatments that the equations used are applicable to arbitrarily large perturbations, and further that the fluid is incompressible, to the extent that the incompressibility condition $\nabla\cdot {\bf v} = 0$ is used as if it had the rank of a transport equation, qualified to be used in place of the closing equation for the pressure. The equations used in the quoted papers are

\begin{equation} 
\frac{\partial \bf v}{\partial t} + ({\bf v} \cdot \nabla ) {\bf v} = -\nabla
\bar{p}
\end{equation}
\begin{equation} 
\nabla\cdot {\bf v} = 0
\end{equation}

where $\bar{p} = p / Nm$, $p$ is the pressure, $N$ the particle number density (assumed to be constant), and $m$ the particle mass.

The Navier--Stokes equation, on the other hand, has received renewed interest through the Millennium Prize endowed by the Clay Mathematics Institute (Fefferman [6]). According to the Clay Institute's judgement, solutions of

\begin{equation} 
\frac{\partial \bf v}{\partial t} + ({\bf v} \cdot \nabla ) {\bf v} = -\nabla
\bar{p} + \mu \, \Delta {\bf v}
\end{equation}

in conjunction with eq.~(2) belong to the seven most important open problems in mathematics. The prescribed task is to find a proof for the existence and smoothness of solutions of [(3),(2)], and the expectation is that thereby progress be made to unlock the secrets hidden in the Navier--Stokes equation. Realizing that the system [(3),(2)] is built on drastic physical simplifications, it is not obvious that a successful accomplishment of the mathematical task would have physical relevance, and indeed we will see that it has not. In (3), $\mu$ is the kinematic viscosity.

It will be the purpose of the present paper to examine the physical consequences of the simplifying assumptions entering into the equations [(1),(2)] and [(3),(2)], respectively. 

\vspace*{5mm} {\large \bf 2. Basic equations}

The mother equation for any transport equation is the continuity equation in phase space, describing the distribution function $f({\bf u}, {\bf r}, t)$, normalized by $\int f \, d^3{\bf u} = N$. Considering a one--constituent neutral fluid without external source or loss terms, we have

\begin{equation} 
\frac{d\, f}{d\, t} + ({\bf u} - {\bf v})\cdot \nabla f + \frac{{\bf F}}{m} \cdot \nabla_u f = \frac{\delta f}{\delta t}
\end{equation}

where $\delta f/\delta t$ symbolically denotes the temporal change of $f$ due
to the action of collisions, and {\bf F} is a force acting on the particles individually. In the given context, {\bf F} represents the intermolecular force. The kinetic equation (4) relates to point particles, and to ideal gases if {\bf F} is set to zero. 

It will be assumed that collisions are sufficiently soft to be non--ionizing. Another assumption will be that the particles have no internal degrees of freedom. The inclusion of internal excitations would mean that separate equations would have to be formulated for every excitation level, with terms describing the transitions between them. Such a treatment would be beyond the scope of the present work, and so we have to take it as given that the particles have only their three translatorial degrees of freedom which explains the occurrence of the factors 3/2 or 2/3 in some of the subsequent relations.

Eq.~(4) will be used to derive a system of transport equations for the moments of the distribution function. Before proceeding, we will distinguish two groups of moments, those originating from the distribution function in phase space, $f$, and those originating from the distribution function in velocity space, $\tilde{f} = f/N $. The macroscopic velocity {\bf v} and the kinetic temperature $T$ belong to the second group,

\begin{equation} 
{\bf v} = \int \tilde{f} \, {\bf u}\, d^3{\bf u}
\end{equation}

\begin{equation} 
T = \frac{m}{3\, K}\int \tilde{f} \, ({\bf u} -{\bf v})^2 d^3{\bf u}
\end{equation}

where $K$ is Boltzmann's constant. A complete set of moments of $f$ can be defined by

\begin{equation} 
M^{(i,j,k)}({\bf r}, t) =  \int f({\bf u}, {\bf r}, t) \, \phi^{(i,j,k)}({\bf u}) \, d^3{\bf u}
\end{equation}
with
\begin{equation} 
\phi^{(i,j,k)}({\bf u}) = m \, (u_x - v_x)^i (u_y - v_y)^j (u_z - v_z)^k
\end{equation}

where $i$, $j$, $k$ are non--negative integers, and $n = i+j+k$ is the order of the moment. Multiplication of the terms in (4) by $\phi^{(i,j,k)}$ with subsequent intergration over velocity space yields, after some elementary mathematical steps,

\begin{eqnarray} 
& & \bigg\{ \frac{dM^{(i,j,k)}}{dt} + M^{(i,j,k)}\Big( \,\nabla\cdot {\bf v} + i\,\frac{\partial v_x}{\partial x} + j\,\frac{\partial v_y}{\partial y} + k\,\frac{\partial v_z}{\partial z} \Big)  - \frac{\delta M^{(i,j,k)}}{\delta t}    \bigg\} \mbox{\hspace{1mm}}  \nonumber \\ +\mbox{\hspace{-6mm}} & & \bigg\{ i\, M^{(i-1,j,k)}\,\Big(\frac{dv_x}{dt}-\frac{F_x}{m} \Big) +  j\, M^{(i,j-1,k)}\,\Big(\frac{dv_y}{dt}-\frac{F_y}{m} \Big) + k\, M^{(i,j,k-1)}\,\Big(\frac{dv_z}{dt}-\frac{F_z}{m} \Big) \bigg\} \nonumber \\
+\mbox{\hspace{-6mm}} & &  \bigg\{ i\,M^{(i-1,j+1,k)}\,\frac{\partial v_x}{\partial y} +  j\,M^{(i+1,j-1,k)}\,\frac{\partial v_y}{\partial x} + j\,M^{(i,j-1,k+1)}\,\frac{\partial v_y}{\partial z} + \nonumber \\
& & \mbox{\hspace{3mm}}  k\,M^{(i,j+1,k-1)}\,\frac{\partial v_z}{\partial y} +  k\,M^{(i+1,j,k-1)}\,\frac{\partial v_z}{\partial x} +  i\,M^{(i-1,j,k+1)}\,\frac{\partial v_x}{\partial z} \bigg\}  \nonumber \\
= \mbox{\hspace{-7.3mm}} & & - \, \bigg\{ \frac{\partial M^{(i+1,j,k)}}{\partial x} + \frac{\partial M^{(i,j+1,k)}}{\partial y} + \frac{\partial M^{(i,j,k+1)}}{\partial z} \bigg\} 
\end{eqnarray}

whereby the entirety of transport equations is concentrated in one
equation. For the conversion of the last left--hand side term in (4) into a moment term it has been necessasry to postulate that the decrease of $f$ is stronger than the increase of $\phi^{(i,j,k)}$ as $u$ goes to infinity.

The terms in (9) have been ordered by setting curly brackets. The first pair of brackets contains the wanted moment of order $n$, the second moments of order $n-1$, the third other moments of order $n$, and the fourth moments of order $n+1$. Eq.~(9) shows that any wanted moment $M^{(i,j,k)}$ is coupled with a finite number of moments of same or lower order, and an infinite number of moments of higher order. This reminds us of the well--known, but often forgotten fact that transport equations do not exist as isolated entities, but only as integral parts within a system of transport equations which have to be extracted from the infinite system on the basis of precisely stated truncation conditions.   

Before making use of (9), we identify the moments of orders $n=0$ to $3$.

For $n=0$: \vspace*{4mm}
\begin{equation} 
M^{(0,0,0)} = Nm
\end{equation}

For $n=1$: \vspace*{4mm}
\begin{equation} 
M^{(1,0,0)} = M^{(0,1,0)} = M^{(0,0,1)} = 0
\end{equation}

For $n=2$: \vspace*{4mm}
\begin{equation} 
M^{(2,0,0)} = p_{xx} \mbox{\hspace{4mm},\hspace{4mm}} M^{(0,2,0)} = p_{yy}
\mbox{\hspace{4mm},\hspace{4mm}} M^{(0,0,2)} = p_{zz}
\end{equation}
\vspace*{-2mm}
\begin{equation} 
M^{(1,1,0)} = p_{xy} = p_{yx}\mbox{\hspace{4mm},\hspace{4mm}} M^{(1,0,1)} =
p_{xz} = p_{zx} \mbox{\hspace{4mm},\hspace{4mm}} M^{(0,1,1)} = p_{yz} = p_{zy}
\end{equation}
\vspace*{-2mm}
\begin{equation} 
T=\frac{1}{3KN}\, (M^{(2,0,0)} + M^{(0,2,0)} + M^{(0,0,2)})
\end{equation}

For $n=3$: \vspace*{4mm}
\begin{equation} 
q_x = \frac{1}{2}\,( M^{(3,0,0)} + M^{(1,2,0)} + M^{(1,0,2)})
\end{equation}
\vspace*{-2mm}
\begin{equation} 
q_y = \frac{1}{2}\,( M^{(2,1,0)} + M^{(0,3,0)} + M^{(0,1,2)})
\end{equation}
\vspace*{-2mm}
\begin{equation} 
q_z = \frac{1}{2}\,( M^{(2,0,1)} + M^{(0,2,1)} + M^{(0,0,3)})
\end{equation}

These are the three components of the heat flux vector {\bf q}.

\vspace*{5mm} {\large \bf 3. Transport equations for the fluid variables $N$, {\bf v} and $T$}

Eq.~(9), together with the moment definitions (10) to (17), will now be used to extract equations for $N$, {\bf v} and $T$, neglecting intermolecular forces. We obtain:

For $n=0$: \vspace*{4mm}
\begin{equation} 
\frac{dN}{dt} = - N \, \nabla \cdot {\bf v}
\end{equation}

For $n=1$: \vspace*{4mm}
\begin{equation} 
\frac{d\bf v}{dt} = - \frac{1}{Nm} \left[\, \nabla \, (NKT)  +  \nabla \cdot (\textsf{p})^{\rm o}\, \right]
\end{equation}

where $(\textsf{p})^{\rm o}$ is the traceless part of the pressure tensor $\textsf{p}$, with elements defined by (12) and (13). 
The term in the first round brackets in (19) corresponds to the kinetic pressure 

\begin{equation} 
p = NKT
\end{equation}

of an ideal gas.

For $n=2$: \vspace*{4mm}
\begin{equation} 
\frac{d T}{d t} = - \frac{2}{3} \, T \, \nabla \cdot {\bf v} - \frac{2}{3KN}\,\left[ \, (\textsf{p})^{\rm o} : \nabla {\bf v} +
\nabla \cdot {\bf q} \, \right]
\end{equation}

where $\nabla {\bf v}$ denotes the dyadic product. 

The reason why we have no collision terms in the transport equations for $N$, {\bf v} and $T$ is because we have assumed that collisions are non--ionizing ($ \delta N/\delta t = 0$), and because there exists no second constituent with which momentum and energy could be exchanged ($ \delta {\bf v}/\delta t = 0$ and $ \delta T/\delta t = 0$). However, the randomizing effect of collisions will play a dominant role in the subsequent treatment of $(\textsf{p})^{\rm o}$ and ${\bf q}$.

There are two reasons why the equations (18) to (21) are not applicable to fluids in general, but only to ideal gases. The first is that intermolecular forces have been ignored. The second is that the whole formulation is built on the idealization that the particles are infinitesimally small, so that a given volume is able to accomodate an infinite number of particles. This becomes apparent from the continuity equation (18) which allows $N$ to become arbitrarily large. 

In order to get an indication of the applicability limits of eqs.~(18) to (21), we make recourse to van der Waal's equation of state, written here in the form

\begin{equation} 
p_t = p_k - p_i \mbox{\hspace*{6mm} with \hspace{3mm}} p_k = \frac{NKT}{1-V_0\, N} \mbox{\hspace*{3mm} and \hspace{3mm}} p_i = \frac{a}{A^2}\, N^2
\end{equation} 

where $p_t$ is the total pressure, $p_k$ the kinetic prassure, and $p_i$ the internal (or cohesive) pressure. The coefficient $a$, for various gases and liquids, is a tabulated quantity, $A$ is the Avogadro number, and $V_0$ is a measure of the volume of a single particle.

We see that the numerator in the expression for $p_k$ represents the kinetic pressure of an ideal gas, whereas the denominator accounts for the fact that the particles have a finite volume, represented by $V_0$. Obviously, for having an ideal gas, the denominator must be close to 1, and thus $NV_0 \ll 1$. The corresponding smallness condition for the internal pressure is $(a/A^2)N^2 \ll NKT$. So we find that the equations (18) to (22) will be applicable if the two conditions  

\begin{equation} 
N \ll \frac{1}{V_0} \mbox{\hspace*{7mm} and \hspace*{7mm}} N \ll \frac{KTA^2}{a}
\end{equation}

are satisfied.

The above transport equations (19) and (21) are open--ended since the moments $(\textsf{p})^{\rm o}$ and ${\bf q}$ establish a link to an infinite number of higher order moments. It will be necessary, therefore, to find criteria allowing to express $(\textsf{p})^{\rm o}$ and ${\bf q}$ in terms of $N$, {\bf v} and $T$, in order to obtain a closed system of transport equations with no other variables than these.  This  task will be addressed in sections 4 and 5 below.

\vspace*{5mm} {\bf \large 4. Approximation 1: Fully isotropized system}

If the dominant role of randomizing collisions goes to the extreme that the system is completely isotropized, the $(\textsf{p})^{\rm o}$ and ${\bf q}$ terms disappear, and we have

\begin{equation} 
\frac{d \bf v}{d t} =  - \frac{1}{Nm} \nabla p = - V^2 \left[\, \frac{\nabla T}{T} + \frac{\nabla N}{N} \, \right]  
\end{equation}

\begin{equation} 
\frac{d T}{d t} = - \frac{2}{3} \, T \, \nabla \cdot {\bf v}
\end{equation}

where $V$ is the thermal velocity ($V^2 = KT/m$). The first part of eq.~(24) is the Euler equation, however with the restriction that $p$ here is not the pressure in a general sense, but the kinetic pressure given by (20). The equations (18), (24), (25) form a closed set, leaving no room for any other equation for the fluid variables $N$, {\bf v} and $T$.

The derivation shows that the Euler equation has a very narrow range of applicability:  Firstly, the fluid must be treatable as an ideal gas, in the sense of conditions (23), and secondly, deviations of the pressure tensor $\textsf{p}$ from $NKT\, \textsf{U}$ ($\textsf{U}$ = unit tensor) must be exceedingly small (c.f.~conditions (26) below).

\vspace*{5mm} {\bf \large 5. Approximation 2: Nearly isotropized system}

With reference to pressure tensor $(\textsf{p})^{\rm o}$, complete isotropization means that $p_{ij} = 0$ ($i \ne j$) and $p_{ii} = p$. The plausible next step then is to go from completely isotropized to almost isotropized by demanding

\begin{equation} 
|p_{ij}| \ll p_{ii} \mbox{\hspace{.6cm} and \hspace{.6cm}} |p_{ii} - p| \ll p
\end{equation}

where $p$ here stands as abbreviation for $p=(1/3)(p_{xx}+p_{yy}+p_{zz})$. The same conditions apply to the Euler equation, but in a more rigorous way, which might be indicated by using the symbol $\lll$ instead of $\ll$.

In order to obtain local transport coefficients, i.e., transport coefficients which are determined entirely by the properties of the substance, independent of the temporal and spatial development of the process, it has to be demanded that the time between collisions, $\tau_c$, is the shortest time, and the travel distance between collisions, $l_c$, the shortest length in the system,

\begin{equation} 
\tau_c \ll \tau \mbox{\hspace{.6cm} and \hspace{.6cm}} l_c \ll l
\end{equation}

where $\tau$ is a characteristic time and $l$ a characteristic length of the process. These conditions exclude extremely rapid and/or extremely short--scaled processes. 

We consider the pressure tensor elements $p_{xx}$ ($i=2, j=k=0$) and $p_{xy}$ ($i=j=1, k=0$). Since $\delta p/\delta t = 0$, the collision term $\delta p_{xx}/\delta t$ in (9) can be altered into $\delta(p_{xx} - p)/\delta t$, and from the first line in (9), making use of (27), we obtain

\begin{equation} 
\frac{\delta(p_{xx} - p)}{\delta t}\, =\, \frac{d(NKT)}{dt} + NKT \left( \nabla\cdot {\bf v} + 2\, \frac{\partial v_x}{\partial x} \right)
\end{equation}

Next, expressing $dN/dt$ by the continuity equation (18), and $dT/dt$ by the leading term in (21) ($dT/dt = -(2/3)T\nabla\cdot {\bf v}$), we obtain

\begin{equation} 
\frac{\delta(p_{xx} - p)}{\delta t}\, =\, NKT \left( 2\, \frac{\partial v_x}{\partial x} - \frac{2}{3}\, \nabla\cdot {\bf v} \right)
\end{equation}

From the first and third brackets in (9):

\begin{equation} 
\frac{\delta p_{xy}}{\delta t}\, = \, NKT \left( \frac{\partial v_x}{\partial y} + \frac{\partial v_y}{\partial x} \right)
\end{equation}

Corresponding expressions are obtained for the other elements of the pressure tensor, and the results can be written in compact form as

\begin{equation} 
\frac{\delta(\textsf{p})^{\rm o}}{\delta t} =  NKT\, \left[\, \nabla {\bf v} + (\nabla {\bf v})^t - \frac{2}{3}\, (\nabla\cdot {\bf v}) \, \textsf{U} \, \right]
\end{equation}

where $(\nabla {\bf v})^t$ denotes the transposed dyadic.

The derivation of a corresponding expression for $\delta{\bf q}/\delta t$ is a rather intricate matter since it will be necessary to go up to moments of order $n=4$. For these we assume

\begin{equation} 
|M_{odd}^{(i,j,k)}| \ll M_{even}^{(i,j,k)} \mbox{\hspace{.6cm} and \hspace{.6cm}} |M_{even}^{(i,j,k)} - M_{maxw}^{(i,j,k)}| \ll M_{maxw}^{(i,j,k)}
\end{equation}

where $M_{even}^{(i,j,k)}$ means that all indices $i$, $j$, $k$ are even, $M_{odd}^{(i,j,k)}$ that not all indices are even, and $M_{maxw}^{(i,j,k)}$ that $M_{even}^{(i,j,k)}$ is approximated by using a local Maxwell distribution, displaced by the macroscopic velocity {\bf v}. The background for these assumptions is that odd moments occur only in the form of a deviation from equilibrium, whereas even moments also exist in equilibrium. 

We consider the $x$--component of {\bf q}. From (9) with (15), (26), (27), (32) and $F_x = 0$, adding the three equations for $M^{(3,0,0)}$, $M^{(1,2,0)}$ and $M^{(1,0,2)}$, we obtain

\begin{equation} 
\frac{\delta q_x}{\delta t} = \frac{5}{2}\, p\,\frac{dv_x}{dt} + \frac{1}{2}\, \frac{\partial}{\partial x} \left( M^{(4,0,0)}_{\it maxw} + M^{(2,2,0)}_{\it maxw} + M^{(2,0,2)}_{\it maxw} \right)
\end{equation}

With the useful relation

\begin{equation} 
M^{(i,j,k)}_{\it maxw} = (i-1)!!\, (j-1)!!\, (k-1)!!\,\, NKT\, V^{n-2} \, \delta^{(i,j,k)}
\end{equation}

(where $\delta^{(i,j,k)} = 1$ if $i$, $j$, $k$ are all even ; $\delta^{(i,j,k)} = 0$ otherwise), the terms in brackets in (33) add up to $5NK^2T^2/m $. Then, approximating $dv_x/dt$ by the leading term in (19), $-(1/Nm)\, \partial(NKT)/\partial x$, and generalizing from $\delta q_x/\delta t$ to $\delta {\bf q}/\delta t$, we obtain

\begin{equation} 
\frac{\delta {\bf q}}{\delta t} = \frac{5}{2} NKV^2 \, \nabla \, T 
\end{equation}

To finalize the derivation, we need a relation between $\delta(\textsf{p})^{\rm o}/\delta t $ and $(\textsf{p})^{\rm o}$, and between $\delta {\bf q} / \delta t$ and {\bf q}. For this, we use the 10--moment relaxation model of Stubbe [7] and have

\begin{equation} 
\frac{\delta(\textsf{p})^{\rm o}}{\delta t} =  - \, 2\, \nu^{(r)} (\textsf{p})^{\rm o}
\end{equation}

where the collision frequency $\nu^{(r)}$ is a measure of the strength of the randomzing action of collisions, quantitatively defined by eqs.~(28c) and (15) of [7]. From (31) and (36)

\begin{equation} 
(\textsf{p})^{\rm o} =  - \,  \eta \left[\, \nabla {\bf v} + (\nabla {\bf v})^t - \frac{2}{3}\, (\nabla\cdot {\bf v}) \, \textsf{U} \, \right]
\end{equation}
where
\begin{equation} 
\eta = \frac{NKT}{2\, \nu^{(r)}}
\end{equation}

is the dynamic viscosity. In a corresponding way, {\bf q} is obtained as

\begin{equation} 
{\bf q} = - \, \kappa \, \nabla \, T
\end{equation}
where
\begin{equation} 
\kappa = \frac{5\,NK\,V^2}{4\, \nu^{(r)}}
\end{equation}

is the heat conductivity. Since $\nu^{(r)}$ is proportional to $N$, both $\eta$ and $\kappa$ are independent of $N$. The viscosity given by (38) agrees fully with the corresponding result provided by the first approximation of the Chapman--Enskog theory (see Chapman and Cowling [8]), whereas in eq.~(40) the factor 5/4 should be replaced by 15/8.

We have now reached our goal to obtain a closed system of equations for the fluid variables $N$, {\bf v} and $T$: The number density $N$ is determined by (18), the macroscopic velocity {\bf v} by (19) in conjunction with (37), and the kinetic temperature $T$ by (21) in conjunction with (37) and (39). The applicability of this system of equations is bound to the conditions (27), (26) and (32). Whereas (27) is easily fulfilled under regular circumstances, (26) and (32) are strongly limiting since they have the character of a hidden linearization.

In order obtain the Navier--Stokes equation from (19) with (20) and (37), we have to assume that $\eta$ is constant (i.e., $\nabla \eta$ sufficiently small). Hereby, (19) becomes

\begin{equation} 
\frac{d \bf v}{d t} =   - \,  V^2 \left[\, \frac{\nabla T}{T} + \frac{\nabla N}{N} \, \right] +\frac{\eta}{N m} \, \nabla \cdot  \left[\, \nabla {\bf v} + (\nabla {\bf v})^t - \frac{2}{3}\, (\nabla\cdot {\bf v}) \, \textsf{U} \, \right]
\end{equation}

It is straightforward to convert (41) into the equivalent form

\begin{equation} 
\frac{d \bf v}{d t} =  - \,  V^2 \left[\, \frac{\nabla T}{T} + \frac{\nabla N}{N} \, \right] + \frac{\eta}{N m} \, \left[\,\Delta {\bf v} + \frac{1}{3}\, \nabla(\nabla\cdot {\bf v}) \,\right]
\end{equation}

which is the Navier--Stokes equation, except that usually the first right--hand side term is written as $-(1/Nm)\nabla p$, enabled by (20).

Similarly, treating $\kappa$ as a constant, (21) in conjunction with (37) and (39) is converted into

\begin{equation} 
\frac{dT}{dt} = -\frac{2}{3}\, T \, \nabla \cdot {\bf v}
+ \frac{2}{3}\, \frac{\eta}{KN}\, \left[ \nabla {\bf v} : [\nabla {\bf v} + (\nabla {\bf v})^t] - \frac{2}{3} (\nabla\cdot {\bf v})^2 \right] + \frac{2}{3}\, \frac{\kappa}{KN}\, \Delta T
\end{equation}

Just as in the case of the Euler system [(18),(19),(21)], the equations (41) (or (42)) and (43), in conjunction with the continuity equation (18), represent an unseparable set, tied together by  joint closure prescriptions. The number of equations matches the number of unknowns, leaving no room for any additional equation for the fluid variables $N$, {\bf v} and $T$. And just as in the case of the Euler system, the Navier--Stokes system [(18),(41),(43)] is valid for ideal gases only. 

The validity of the momentum equation (41) and the thermal equation (43) is strongly restricted by the conditions (26) and (32). The consequence of these conditions is that the viscosity and heat conduction terms have to be small in comparison with the leading terms, i.e., the first terms each on the right--hand sides of (41) and (43). This offers an easy way to check mathematical solutions of the system [(18),(41),(43)] for physical validity.

\vspace*{5mm} {\bf \large 6. Energy equations}

As a useful addition, we convert (41) and (43) into energy equations. The kinetic energy density of an ensemble of particles due to its bulk motion is given by

\begin{equation} 
w_K =  \frac{1}{2}\, m \int f \, {\bf v}^2 d^3 {\bf u} =  \frac{1}{2}N m{\bf v}^2
\end{equation}

and the internal energy density by

\begin{equation} 
w_I = \frac{1}{2}\, m \int f \, ({\bf u} -{\bf v})^2 d^3{\bf u}  = \frac{3}{2}\, NKT  = \frac{3}{2}\, p
\end{equation}

From (19), multiplied by {\bf v}, and (21), the energy equations

\begin{equation} 
\frac{\partial w_K}{\partial t} + \nabla \cdot ( w_K \, {\bf v}) = - \, {\bf v}\cdot \left[\, \nabla p + \nabla \cdot (\textsf{p})^{\rm o} \, \right]
\end{equation}

\begin{equation} 
\frac{\partial w_I}{\partial t} + \nabla \cdot (w_I \, {\bf v}) = - \, \left[\, p \, \nabla \cdot {\bf v} +(\textsf{p})^{\rm o} : \nabla {\bf v} \right]  - \nabla \cdot {\bf q}
\end{equation}

are obtained. With the identity $\; {\bf v}\cdot [\,  \nabla \cdot (\textsf{p})^{\rm o}] + (\textsf{p})^{\rm o} : \nabla {\bf v} = \nabla\cdot [ (\textsf{p})^{\rm o}\cdot {\bf v}]  $ , the sum of $w_K$ and $w_I$, $w = w_K + w_I$, is given by

\begin{equation} 
\frac{\partial w}{\partial t} + \nabla \cdot (w \, {\bf v}) = - \nabla \cdot (p \, {\bf v}) -  \nabla\cdot [ (\textsf{p})^{\rm o}\cdot {\bf v}]  - \nabla \cdot {\bf q}
\end{equation}

implying that the total energy $\int w\, d^3{\bf r}$ inside a closed solid surface is constant. 

Insertion of (37) and (39) in (46) to (48) yields

\begin{equation} 
\frac{\partial w_K}{\partial t} + \nabla \cdot ( w_K \, {\bf v}) = - \, {\bf v}\cdot \nabla p +  \eta \, {\bf v}\cdot  \left( \nabla \cdot \left[\, \nabla {\bf v} + (\nabla {\bf v})^t - \frac{2}{3}\, (\nabla\cdot {\bf v}) \, \textsf{U} \, \right] \right)
\end{equation}

\begin{equation} 
\frac{\partial w_I}{\partial t} + \nabla \cdot (w_I \, {\bf v}) = - \,   p \, \nabla \cdot {\bf v} + \eta \, \left[ \nabla {\bf v} : [\nabla {\bf v} + (\nabla {\bf v})^t] - \frac{2}{3} (\nabla\cdot {\bf v})^2 \right] + \kappa \, \Delta T
\end{equation}

\begin{equation} 
\frac{\partial w}{\partial t} + \nabla \cdot (w \, {\bf v}) = - \nabla \cdot (p \, {\bf v}) +  \eta \, \nabla \cdot  \left( {\bf v} \cdot \left[\, \nabla {\bf v} + (\nabla {\bf v})^t - \frac{2}{3}\, (\nabla\cdot {\bf v}) \, \textsf{U} \, \right] \right) + \kappa \, \Delta T
\end{equation}
 
Any mathematical solution of the system of transport equations has to satisfy the energy equation (51) in order to be regarded as physically valid.

\vspace*{5mm} {\bf \large 7. Incompressible flow within the frame of the Navier--Stokes equation}

As we have seen, the Navier--Stokes equation does not exist as an isolated equation, but only as a part of a system of equations, based on well--defined joint applicability conditions, one of these being that the fluid under consideration is an ideal gas. Since an ideal gas is the least incompressible among all conceivable fluids, it appears pointless do devote much attention to this special case which, in the end, may have no relation to any physical reality.   

The incompressibility condition $\nabla\cdot {\bf v} =0 $ is commonly used in combination with $\nabla N = 0$,

\begin{equation} 
\nabla\cdot {\bf v} =0 \mbox{\hspace*{5mm} and \hspace*{5mm}} \nabla N = 0
\end{equation}

implying that $N({\bf r},t)$ is constant and the continuity equation thus not needed.

The conditions (52) are frequently used, but seldom backed by a solid foundation. A typical example of a quasi--justification, taken from Landau and Lifshitz [9], reads as follows: ``In a great many of cases of the flow of liquids, and also of gases, their density may be supposed invariable, i.e.~constant throughout the volume of the fluid and throughout its motion. In other words, there is no noticeable compression or expansion of the fluid in such cases. We then speak of incompressible flow.'' But why should one not as well write ``In a great many of cases, the temperature may be supposed invariable. We then speak of isothermal flow''? Both statements appear  equally arbitrary.

Despite these reservations, we will examine to what extent (52) will be able to simplify matters. In the literature, the conditions (52) are commonly used twofold, firstly to simplify the Navier--Stokes equation, and secondly to discard the temperature equation (43) and use $\nabla\cdot {\bf v} =0$ instead. The system [(3),(2)] is an example: The Navier--Stokes equation there is written in the simplified form (3), and eq.~(2) stands where an equation for  $\bar{p}$ should stand.

The common way to employ the condition $\nabla\cdot {\bf v} =0 $ in the Navier--Stokes equation (42) is to ignore the $\nabla(\nabla\cdot {\bf v})$ term,

\begin{equation} 
\frac{d \bf v}{d t} =  - \,  V^2 \, \frac{ \nabla T}{T}\, +\, \frac{\eta}{N m} \, \,\Delta {\bf v}
\end{equation}

Realizing that $\Delta {\bf v} =  \nabla(\nabla\cdot {\bf v}) - \nabla \times (\nabla \times {\bf v} )$, a more appropriate formulation would be

\begin{equation} 
\frac{d \bf v}{d t} =  - \,  V^2 \, \frac{\nabla T}{T}\, - \, \frac{\eta}{N m} \, \nabla \times (\nabla \times {\bf v} )
\end{equation}

A third approximation is obtained from (41),

\begin{equation} 
\frac{d \bf v}{d t} =  - \,  V^2 \, \frac{\nabla T}{T}\, + \, \frac{\eta}{N m} \, \nabla \cdot  [\, \nabla {\bf v} + (\nabla {\bf v})^t ]
\end{equation}

In order to decide which of these three simplified momentum equations should be used, we go back to the origin, eq.~(9). We see that the only occurrence of $\nabla \cdot {\bf v}$ there is in the first line. Setting this term to zero and following the derivation through to the Navier--Stokes approximation, we arrive at (55) which, therefore, is to be regarded as the appropriate one among the three versions.

The thermal equation (43) with $\nabla \cdot {\bf v} = 0$  becomes

\begin{equation} 
\frac{dT}{dt} =  \frac{2}{3}\, \frac{\eta}{KN}\,  \nabla {\bf v} : [\nabla {\bf v} + (\nabla {\bf v})^t]  + \frac{2}{3}\, \frac{\kappa}{KN}\, \Delta T   
\end{equation}

The simplified equations (55) and (56) form a closed set, with $N$ treated as an externally given constant. This is as far as one can get by using the assumptions (52) for a simplification of the momentum and thermal equation.

The next step in the chain of approximations will be to discard (56) and use (2) instead. To see what the physical implications of this  procedure will be, we go back to the thermal energy equation (47), write out the terms in squared brackets in component form, and define: 

\begin{equation} 
\dot{w}_{I\leftrightarrow P} = - \, p \, \nabla \cdot {\bf v} - (p_{xx}-p)\frac{\partial v_x}{\partial x} - \, (p_{yy}-p)\frac{\partial v_y}{\partial y} - \, (p_{zz}-p)\frac{\partial v_z}{\partial z}
\end{equation}

\begin{equation} 
\dot{w}_{K\rightarrow I} = - \, p_{xy}\left(\frac{\partial v_x}{\partial y} + \frac{\partial v_y}{\partial x} \right) - \, p_{yz}\left(\frac{\partial v_y}{\partial z} + \frac{\partial v_z}{\partial y} \right) - \,
p_{xz}\left(\frac{\partial v_z}{\partial x} + \frac{\partial v_x}{\partial z} \right)
\end{equation}

\begin{equation} 
\dot{w}_{I\leftrightarrow I} = \kappa \, \Delta T
\end{equation}
 
Thereby, eq.~(47) is brought into the concise form

\begin{equation} 
\frac{\partial w_I}{\partial t} + \nabla \cdot  (w_I \, {\bf v}) = \dot{w}_{I\leftrightarrow P} + \dot{w}_{K\rightarrow I} + \dot{w}_{I\leftrightarrow I} 
\end{equation}

The meaning of the energy exchange terms $\dot{w}_{I\leftrightarrow P}$ and $\dot{w}_{I\leftrightarrow I}$ is obvious:  $\dot{w}_{I\leftrightarrow P}$ describes the mutual conversion of potential and internal energy, due to compression or dilatation, and $\dot{w}_{I\leftrightarrow I}$ describes the local redistribution of internal energy due to conduction of heat. With this given, the remaining term in (60), $\dot{w}_{K\rightarrow I}$, is left to describe the one--sided conversion of kinetic into internal energy due to internal friction. This interpretation is supported by the fact that $\dot{w}_{K\rightarrow I}$ turns out to be an unconditionally positive quantity, as seen by insertion of (37) in (58):

\begin{equation} 
\dot{w}_{K\rightarrow I} =  \eta \left[ \left( \frac{\partial v_x}{\partial y} 
+ \frac{\partial v_y}{\partial x} \right)^{\mbox{\hspace*{-1.2mm}}2} + \left( \frac{\partial v_y}{\partial z} + \frac{\partial v_z}{\partial y} \right)^{\mbox{\hspace*{-1.2mm}}2} +\left( \frac{\partial v_z}{\partial x} + \frac{\partial v_x}{\partial z} \right)^{\mbox{\hspace*{-1.2mm}}2} \, \right]
\end{equation}

The physical function of this term is to irreversibly transfer energy from the kinetic into the thermal energy reservoir. In a complete physical system, the gain term $\dot{w}_{K\rightarrow I}$ in (60) will be balanced by a corresponding loss term contained in the kinetic energy equation, and both will add up to zero. However, the system [(55),(2)] is not complete: The thermal energy equation is removed and replaced by the non--thermal equation $\nabla \cdot {\bf v} = 0 $, with the consequence that kinetic energy will be lost without appearing elsewhere. The total energy of the system will not be conserved. 

While $\dot{w}_{K\rightarrow I}$ is not affected by the conditions (52), $\dot{w}_{I\leftrightarrow P}$ is. Insertion of (37) in (57) with (52) yields

\begin{equation} 
\dot{w}_{I\leftrightarrow P} = 2\, \eta \left[ \left( \frac{\partial v_x}{\partial x}\right)^{\mbox{\hspace*{-1.2mm}}2} + \left( \frac{\partial v_x}{\partial x}\right)^{\mbox{\hspace*{-1.2mm}}2}  + \left( \frac{\partial v_x}{\partial x}\right)^{\mbox{\hspace*{-1.2mm}}2} \, \right]
\end{equation}

The conditions (52) are meant to exclude compressions or dilatations, implying that $\dot{w}_{I\leftrightarrow P}$
should vanish, yet we see from (62) that $\dot{w}_{I\leftrightarrow P}$ in fact does not fully disappear. The remaining part (62), being entirely positive, acts as if compressions are possible, but not dilatations. This is another unacceptable consequence.

A third inconsistency is that use of the system [(55),(2)] includes the possibility of a local accumulation or depletion of thermal energy, but contains no mechanism for a redistribution by means of heat conduction.

We conclude that the replacement of the thermal equation (56) by the incompressibility condition $\nabla \cdot {\bf v} = 0$ has unacceptable consequences, even in the case that this condition as such might be justifiable. All that this condition can possibly do is to bring the full system of transport equations into the simplified form [(55),(56)]. As a consequence, equations [(3),(2)] will not be able to describe a real physical situation. Their solutions will be of sole mathematical interest. 

\vspace*{5mm} {\bf \large 8. Incompressible flow within the frame of the Euler equation}

Repeating from section 4, the Euler system is given by \setcounter{equation}{23}

\begin{equation} 
\frac{\partial \bf v}{\partial t}  + ({\bf v} \cdot \nabla ){\bf v} =   - V^2 \left[\, \frac{\nabla T}{T} + \frac{\nabla N}{N} \, \right] = - \frac{1}{Nm} \nabla p
\end{equation}

\begin{equation} 
\frac{\partial T}{\partial t} + {\bf v} \cdot \nabla T = - \frac{2}{3} \, T \, \nabla \cdot {\bf v}
\end{equation}

\setcounter{equation}{62} with $p = NKT$ and $N$ to be determined by the continuity equation (18). 

Simplifying [(24),(25)] by means of the incompressibility conditions (52), we have

\begin{equation} 
\frac{\partial \bf v}{\partial t}  + ({\bf v} \cdot \nabla ){\bf v}  =   - V^2 \, \frac{\nabla T}{T} 
\end{equation}

\begin{equation} 
\frac{\partial T}{\partial t} + {\bf v} \cdot \nabla T = 0
\end{equation}

These equations are now independent of $N$, and the temperature gradient is the only driving agent. Again, we have a closed set of equations, describing the fluid variables {\bf v} and $T$, and again there is no place for any additional equation. 

The fact that these equations are based on the assumption $\nabla \cdot {\bf v} = 0$ does not imply that their solutions will actually confirm the validity of this assumption. Therefore, if \{${\bf v}_1({\bf r},t),T_1({\bf r},t)$\} is a mathematical solution of [(63),(64)], it will be necessary to show a posteriori that $\nabla \cdot {\bf v}_1 \approx 0$ for all {\bf r} and $t$. 

In this situation it may be tempting to enforce incompressibility by using eq.~(2), $\nabla \cdot {\bf v} = 0$, in place of (64). Although this is common practice, as exemplified by [(1),(2)], it does not make things easier, because now it has to be shown that any solution \{${\bf v}_2({\bf r},t),T_2({\bf r},t)$\} of [(63),(2)] obeys the temperature equation (64). This equation does not lose its existence only because the fluid is assumed to be incompressible. 

In the end we see that the full Euler system [(24),(25)] is easier to handle than its simplified versions [(63),(64)] or [(63),(2)]. Incompressibility is not a property that can be enforced; if existing, it will have to result self--consistently from the full equations.

The arguments presented here apply correspondingly to the Navier--Stokes system, in addition to those given in section 7.

\vspace*{5mm} {\bf \large 9. Summary}

It has been the purpose of the present paper to work out the applicability limits of the Euler and Navier--Stokes equations, and to examine the consequences of the frequently used incompressibi\-lity assumption. To this end, the Euler and Navier--Stokes equations have been rederived, starting at the kinetic equation for the distribution function in phase space. 

It follows from this derivation that the applicability of the Euler and Navier--Stokes equations is bound to the following conditions:

(a) The localization condition (27), securing that the transport coefficients are matter constants, and excluding extremely rapid and/or short--scaled processes.  

(b) The isotropization conditions (26) and (32). These have the character of a hidden linearization, implying that only solutions desribing small deviations from equilibrium are physically valid. 

(c) The rarefaction conditions (23). These exclude the applicability of the Euler and Navier--Stokes equations to dense gases and liquids. For extremely rarefied gases, a conflict may arise between conditions (a) and (c), with the consequence that fluid theory becomes inapplicable altogether.

(d) The energy condition, consisting in the requirement that the energy equation (51) be satisfied, corresponding to the conservation of the total energy in a closed system. If mathematical solutions are obtained that do not meet this requirement, it will be pointless to search for physical explanations, there are none. 

The conditions (a) to (d) set a dividing line between physically valid and invalid solutions, and this dividing line must be observed when hydrodynamic equations are used as material for mathematical studies. 

Further, it follows that the system of equations comprising the Euler or Navier--Stokes equation forms a complete set for the fluid variables $N$, {\bf v} and $T$, leaving no place for any other equation, like the relation $\nabla \cdot {\bf v} = 0$ used to enforce incompressibility. Consequently, the equations  [(1),(2)] or [(3),(2)] are without physical foundation and must be left to purely mathematical interest.

\vspace*{5mm} {\bf \large Acknowledgment}

Fruitful discussions with Bernd Inhester at an early stage of this work are acknowledged. Thanks are due to Cedric Villani for directing my attention to the mathematical work quoted in the introduction, which was the stimululus for the present work.

\vspace*{1.5cm}

\centerline {\bf References}
\vspace*{3mm}

[1] V.~Scheffer, {\it An inviscid flow with compact support in spacetime},
J.~Geom.~Analysis {\bf 3} (1993), 343--401.

[2] A.~Shnirelman, {\it On the nonuniqueness of weak solutions of the Euler
equation}, Comm.~Pure \& Appl.~Math. {\bf 50} (1997), 1260--1286.

[3] P.~Constantin, E.~Weinan and E.~Titi, {\it Onsager´s conjecture on the
energy conservation for solutiuons of Euler´s equation}, Comm.~Math.~Phys. {\bf
165} (1994), 207--209.

[4] A.~Cheskidov, P.~Constantin, S.~Friedlander and R.~Shvydkoy, {\it Energy
conservation and Onsager´s conjecture for the Euler equations}, Nonlinearity
{\bf 21} (2008), 1233--1252.

[5] C.~De Lellis and L.~Szekelyhidi, {\it Dissipative continuous Euler flows},
Inventiones Mathematicae {\bf 193}, Issue 2 (2013), 377--407.

[6] C.~Fefferman, {\it Existence and smoothness of the Navier--Stokes equation}, http://www.claymath.org/
millennium/Navier-Stokes\_Equations/navierstokes.pdf

[7] P.~Stubbe, {\it A new collisional relaxation model for small deviations from equilibrium}, J.~Plasma Physics {\bf 38}, Part 1 (1987), 95--116.

[8] S.~Chapman and T.~Cowling, {\it The Mathematical Theory of Non--Uniform
Gases}, Cambridge University Press (1970).

[9] L.~Landau and E.~Lifshitz, {\it Course of Theoretical Physics, Vol.~6: Fluid Mechanics}, Pergamon Press (1987).

\end{document}